# Tunable thermal conductivity of sustainable geopolymers by Si/Al ratio and moisture content: insights from atomistic simulations

*Wenkai Liu and Shenghong Ju**

China-UK Low Carbon College, Shanghai Jiao Tong University, Shanghai, 201306, China.

ABSTRACT. In this work, the effects of Si/Al ratio and moisture content on thermal transport in sustainable geopolymers has been comprehensively investigated by using the molecular dynamics simulation. The thermal conductivity of geopolymer systems increases with the increase of Si/Al ratio, and the phonon vibration frequency region which plays a major role in the main increase of its thermal conductivity is 8-25 THz, while the rest of the frequency interval contribute less. With the increase of moisture content, the thermal conductivity of geopolymer systems decreases at first, then increases and finally tends to be stable, which is contrary to the changing trend of porosity of the system. This is mainly because the existence of pores will lead to phonon scattering during thermal transport, which in turn affects the thermal conductivity of the system. When the moisture content is 5%, the thermal conductivity reaches a minimum value of about 1.103 W/(m·K), which is 40.2% lower than the thermal conductivity of the system without water molecule. This work will help to enhance the physical level understanding of the relationship between the geopolymer structures and thermal transport properties.

KEYWORDS. Sustainable geopolymers, Non-equilibrium molecular dynamics, Moisture content, Si/Al ratio, Thermal transport

## 1. INTRODUCTION

Geopolymers are a kind of inorganic polymer material formed by polycondensation reaction of raw materials containing aluminosilicate and alkaline activating solution[1-3]. The source materials of synthetic geopolymers are very extensive, including metakaolin, fly ash, ground granulated blast furnace slag, red mud, rice husk ash, mine tailings and so on[4-6]. By utilizing these waste materials, it is possible to reduce waste disposal and treatment costs while adding new value and applications to them. This promotes the concept of resource circularity and sustainable utilization[7]. In addition, geopolymers preparation process results in significantly lower carbon emissions compared to the traditional cement preparation process, with reductions exceeding 80%[8]. The traditional cement production process requires high temperature calcination, which is a carbon-intensive process that releases a large amount of carbon dioxide and has a negative impact on climate change[9]. In contrast, geopolymers preparation process involves chemical reactions at lower temperature, thus reducing energy use and associated carbon emissions, and the carbon footprint can be further reduced by selecting alkaline activators[10-11]. At the same time, geopolymers also have many other advantages, such as excellent mechanical properties, fire resistance, chemical corrosion resistance and so on[12-14]. In recent years, They have received increasing research attention and have been widely used in construction[15], chemical industry[16], aerospace[17] and many other fields.

There are many factors that affect the properties of geopolymer materials, such as Si/Al ratio, moisture content, activators, curing conditions and so on[18-20]. Davidovits[21] reported that the structure of geopolymer materials is related with the Si/Al ratio. When the Si/Al ratio in the source material is 1, 2 and 3, the structural of geopolymer will be Poly(sialate) type (-Si-O-Al-O-), Poly(sialate-siloxo) type (-Si-O-Al-O-Si-O-), and Poly(sialate-disiloxo) type (-Si-O-Al-O-Si-O-Si-O-), respectively. Sadat et al.[22] show that the Si/Al ratio significantly affect the existence of non-bridging oxygen atoms, penta coordinated aluminum atoms and edge-sharing aluminum tetrahedra in geopolymers, and then have a significant impact on its failure mechanism and strength. Ahmed et al.[23] found that the thermal conductivity of geopolymer tends to increase with the increase of Si/Al ratio, but its mechanism is not explained. In addition, many studies have shown that most of the water molecules formed in the process of dehydration and condensation of geopolymers will evaporate[24-27]. However, a small amount of water molecules will still exist in the aluminosilicate network, becoming a part of its structural composition and exerting a significant influence on its mechanical and thermal properties[28-30]. Sadat et al.[31] found that for a given Si/Al ratio, the ultimate tensile strength decreases with the increase of moisture content, but the elastic modulus remains unchanged. This is mainly due to the fact that the diffusion of sodium ions increases obviously with the increase of moisture content, thus accelerating the instability of aluminum tetrahedron. Singh[32] found that the thermal conductivity of fly ash-based geopolymer decreases with the decrease of density, and he analyzed that this may be caused by water molecules in the structure. However, the internal mechanism of how water molecules affect the thermal conductivity of geopolymer is still unclear, and it is difficult to solve this problem by traditional experimental method. Molecular dynamics (MD) simulation method provides effective computational method to reveal the microscopic behaviors and properties of materials by detailed atomic and molecular levels information[33-34], making it crucial for understanding molecular mechanisms in geopolymers[35].

Based on the above background, MD simulation has been employed in this work to investigate the effect of different Si/Al ratio and moisture content on the thermal conductivity of geopolymer. In addition, the heat transport mechanism will be analyzed by many analytical methods such as phonon density of states, phonon participation rate and spectral thermal conductivity. The aim of work is to help to enhance the physical level understanding of the relationship between geopolymer structures and heat transport properties, and provide certain guiding significance for future multi-scale simulation on geopolymer composites.

## 2. COMPUTATIONAL DETIALS

### 2.1. Atomic Modeling

In order to study the effect of Si/Al ratio and moisture content on the thermal conductivity of geopolymer systems, systems with Si/Al ratio of 1, 2 and 3 were constructed, respectively. This range encompasses the primary structural types of geopolymers widely studied in the field[21, 31, 36-37]. In addition, systems with Si/Al ratio of 2 and moisture content of 1.25%, 2.5%, 3.75%, 5%, 6.25%, and 7.5% were also constructed (as shown in Table 1). It is worth pointing out that we restricted our study to sodium-based geopolymer with a Na/Al ratio equal to 1 to ensure the system charge balance. According to our previous research work[38], we found that the geopolymer system has no obvious size effect, which means that we do not need to consider the impact of the system size on the calculation results. Therefore, we set the size of simulation systems to 80 Å × 40 Å × 40 Å uniformly.

**Table 1.** Simulated systems parameters

| Name | Si/Al ratio | moisture content (%) | Number of Atoms |
|------|-------------|----------------------|-----------------|
| NAS1 | 1 | 0 | 9464 |
| NAS2 | 2 | 0 | 9500 |
| NAS3 | 3 | 0 | 9516 |
| NASH1.25 | 2 | 1.25 | 9905 |
| NASH2.5 | 2 | 2.50 | 10319 |
| NASH3.75 | 2 | 3.75 | 10745 |
| NASH5 | 2 | 5.00 | 11183 |
| NASH6.25 | 2 | 6.25 | 11630 |
| NASH7.5 | 2 | 7.50 | 12092 |

Since the construction process is similar for different systems, a system with Si/Al ratio of 2 and moisture content of 5% is taken as an example to illustrate the construction process. During the atomistic generation process, the monomer structure of Poly(sialate-siloxo) (-Si-O-Al-O-Si-O-) and sodium atom was first generated. Then, a certain content of monomers were randomly added into a cubic box (80 Å × 40 Å × 40 Å) and the tolerance value of 2.0 Å was set to avoid atomistic overlapping. The system was set up with periodic boundary conditions in all three directions, and was then relaxed under NVT ensemble for 200 ps at 300 K with time step of 1 fs. The temperature was then increased to 4000 K and subsequently decreased to 300 K at a rate of 5 K/ps to ensure the generation of amorphous network structure[30]. The system was further equilibrated for 500 ps under NPT ensemble, followed by a NVT ensemble running for 500 ps at 300 K, producing a stable NAS2 structure (as shown in Figure 1a). And with this structure as a starting point, a certain amount (5wt%) of water molecules were randomly added to it, and the temperature was increased to 800 K and then decreased to 300 K at a rate of 5K/ps. Finally, the system was sequentially equilibrated for 500 ps under NPT and NVT ensembles, respectively, to obtain a stable NASH5 structure (as shown in Figure 1b). And in order to reduce the statistical error, three different initial structures were generated for all systems, and the standard error of the results was calculated as the error bar.

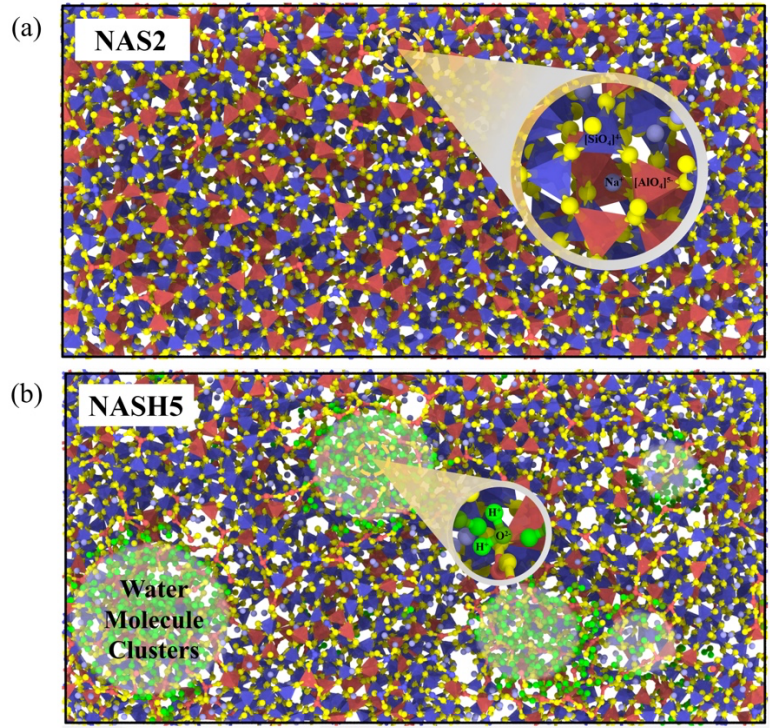

**Figure 1.** (a) The structure diagram of NAS2 system. (b) The structure diagram of NASH5 system (The shaded area indicates water molecule clusters region).

### 2.2. Interatomic potential

In MD simulation, the interatomic potential is crucial to accurately describe the atomistic interactions. In this work, we chose the interatomic potential that have been verified earlier for geopolymer systems with similar composition, including zeolites[39], metakaolin[40] and geopolymers[31, 38]. In order to facilitate the distinction, we mark the oxygen atom in geopolymer structure as $O_s$, the oxygen atom in water molecule as $O_w$, and the hydrogen atom in water molecule as $H_w$. The expression for the interaction potential between M-$O_s$ (M=Si, Al, Na, $O_s$) is[31],

$$V(r) = \frac{Cq_iq_j}{\epsilon r} + A\,exp\left(-\frac{r}{\rho}\right) - \frac{C}{r^6} + \frac{D}{r^{12}} \quad r < r_c, \tag{1}$$

where the first term is the long-range Coulomb interaction in which C is the energy conversion constant, $q_i$ and $q_j$ are the charges of atom $i$ and $j$, $\epsilon$ is the dielectric constant, which is the default value 1.0, $r$ is the distance between the two atoms, $r_c$ is the cutoff radius set as 12 Å. The second and third terms are Buckingham interactions. Usually, the simulation system will be collapsed under the Buckingham potential at high temperature, which is mainly due to that the value of $V(r)$ is negative infinity when $r$ is infinitely small. In this work, we add the fourth repulsive term, which is widely used in other work[41-43]. The detail parameters setting of $q_i$, $q_j$, $A$, $\varrho$, $C$, $D$ are shown in Table 2. The expression for the interaction potential between Si-O$_w$ and Al-O$_w$ is[39],

$$V(r) = \frac{Cq_iq_j}{\epsilon r} + A\,exp\left(-\frac{r}{\rho}\right) + \left(\frac{\sigma}{r}\right)^{12} \quad r < r_c, \tag{2}$$

The detail parameters setting of $q_i$, $q_j$, $A$, $\varrho$, $\sigma$ are shown in Table 2. The interaction between O$_s$-O$_w$ and O$_w$-O$_w$ is the Lennard-Jones potential function, and its expression is[39, 44]:

$$V(r) = 4\varepsilon\left[\left(\frac{\sigma}{r}\right)^{12} - \left(\frac{\sigma}{r}\right)^6\right] \quad r < r_c, \tag{3}$$

The detail parameters setting of $\varepsilon$, $\sigma$ are shown in Table 2. In addition, the interaction in SPC/Fw[44] water molecule model include bond interaction and angle interaction, and the expressions are respectively,

$$V(r) = \frac{1}{2}k_a(r - r_0)^2, \tag{4}$$

$$V(r) = \frac{1}{2}k_b(\theta - \theta_0)^2, \tag{5}$$

where $r_0$ is the equilibrium bond length, $\theta_0$ is the equilibrium bond angle. The detail parameters setting of $k_a, r_0, k_b, \theta_0$ are shown in Table 2. All simulations in this study were conducted by using LAMMPS[45].

**Table 2.** Interatomic potential parameters[31, 40]

| Atoms | Charge | Atoms | Charge |
|---|---|---|---|
| Si | +2.4 | $O_s$ | -1.2 |
| Al | +1.8 | $O_w$ | -0.82 |
| Na | +0.6 | $H_w$ | +0.41 |
| Buckingham potential parameters | | | |
| Atomic pair | $A$(eV) | $\rho$(Å) | $C$(eV·Å$^6$) | $D$(eV·Å$^{12}$) |
| Si-$O_s$ | 13702.9050 | 0.193817 | 54.681 | 2.5 |
| Al-$O_s$ | 12201.4170 | 0.195628 | 31.997 | 1 |
| Na-$O_s$ | 2755.0323 | 0.258583 | 33.831 | 5 |
| Na-$O_w$ | 2616.2137 | 0.258583 | 33.831 | 5 |
| $O_s$-$O_s$ | 2029.2233 | 0.343645 | 192.58 | 150 |
| $O_s$-$H_w$ | 100 | 0.25 | 0 | 0 |
| Si-$O_w$ and Al-$O_w$ potential parameters | | | |
| Atomic pair | $A$(eV) | $\rho$(Å) | $\sigma$(eV$^{-12}$·Å) |
| Si-$O_w$ | 0.5883 | 2.7561 | 1.225 |
| Al-$O_w$ | 4.7788 | 2.5235 | 1.223 |
| Lennard-Jones potential parameters | | | |
| Atomic pair | $\varepsilon$(eV) | $\sigma$(Å) |
| $O_s$-$O_w$ | 0.024309 | 2.4952 |
| $O_w$-$O_w$ | 0.006735 | 3.1690 |
| Bond potential parameters | | | |
| Bond | $k_a$(eV/Å$^2$) | $\sigma$(Å) |
| $O_w$-$H_w$ | 45.93 | 1.012 |

| Angle potential parameters | | |
|---|---|---|
| Angle | $k_b$(eV/rad$^2$) | $\theta_0$ |
| $H_w$-$O_w$-$H_w$ | 3.29136 | 113.24 |

## 2.3. Analytical methods

### 2.3.1. Non-equilibrium molecular dynamics

The thermal conductivity of geopolymer systems was calculated by using the non-equilibrium molecular dynamics method (NEMD)[46]. The initial structures were first equilibrated for 400 ps under NVT ensemble, and the temperature was maintained at 300 K using the Nosé-Hoover heat bath[47-48]. Then, the system was further equilibrated under an NVE ensemble. In order to obtain the temperature distribution, the system was divided into N (N=40) blocks along the heat flux direction, and the temperature of the heat source (the twenty-first block) and heat sink (the first block) were maintained at 340 K and 260 K, respectively, using the Langevin heat bath. The total simulation time was set as 2 ns to make the temperature distribution reach a steady state. The heat flux $J$ can be obtained by calculating the rate ($dE/dt$) of adding or removing energy to the heat sink or source regions, respectively, which is expressed as,

$$J = \frac{dE/dt}{A},$$ (6)

where $A$ is the cross-sectional area in the direction perpendicular to the heat flux. In the statistical stage, we sample the accumulated energy exchange between the system and the thermal baths (as shown in Figure 2a) and the local temperatures (as shown in Figure 2b) for the last 1 ns. The thermal conductivity ($\kappa$) of the system can be obtained by,

$$k = \frac{J}{dT/dx}.$$ (7)

### 2.3.2. Phonon density of states calculation

The phonon density of states (PDOS) is a useful method to characterize phonon activities of materials[49]. In MD, the PDOS is obtained by calculating the velocity autocorrelation function of atoms in the system and then performing Fourier transform[50],

$$PDOS(\omega) = \int_{-\infty}^{\infty} e^{i\omega t} \langle \sum_{j=1}^{N} v_j(t) \cdot v_j(0) \rangle dt, \tag{8}$$

where $v_j(t), v_j(0)$ are the velocities of atom $j$ at moments $t$ and $0$, respectively, $\omega$ is the vibrational frequency, $N$ is the total number of atoms in the system, $PDOS(\omega)$ is the PDOS at the vibrational frequency $\omega$.

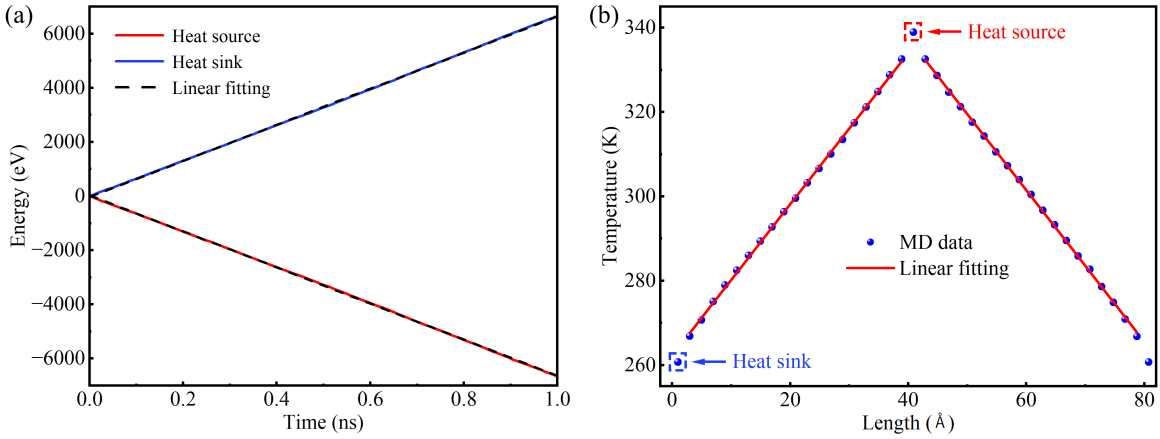

**Figure 2.** (a) The accumulated energy of the Langevin heat bath (heat source and heat sink). (b) Steady-state mean temperature distribution curve for the geopolymer system (NAS2).

### 2.3.3. Phonon participation ratio calculation

The phonon participation ratio (PPR) is another effective method to gain insight into phonon activities information[51], which can be obtained directly from MD simulation without lattice dynamics calculation, greatly improving the computational efficiency. Its expression is

$$PPR(\omega) = \frac{1}{N} \frac{(\sum_i PDOS_i(\omega)^2)^2}{\sum_i PDOS_i(\omega)^4},$$ (9)

where $N$ is the total number of atoms in the system; $PDOS_i(\omega)$ is the PDOS of atom $i$ at the vibrational frequency $\omega$.

### 2.3.4. Spectral thermal conductivity calculation

The spectral heat current (SHC) describes the contribution of phonons with different frequency to heat flux, which is obtained by the calculation of the force and velocity correlation functions near the virtual interface between heat source and sink regions[52-53]. The spectral heat current $q_{i \to j}(\omega)$ between atoms $i$ and $j$ located at different sides of the virtual interface is calculated as,

$$q_{i \to j}(\omega) = -\frac{2}{t_{simu}\omega} \sum_{\alpha,\beta \in \{x,y,z\}} Im\langle \hat{v}_i^{\alpha}(\omega)^* K_{ij}^{\alpha\beta} \hat{v}_j^{\beta}(\omega)\rangle,$$ (10)

where $t_{simu}$ is the simulation time, $\omega$ is the angular frequency, $K_{ij}^{\alpha\beta}$ is the force constant matrix which is calculated by,

$$K_{ij}^{\alpha\beta} = \frac{\partial^2 U}{\partial u_i^{\alpha} \partial u_j^{\beta}}|_{u=0},$$ (11)

where $u_i^{\alpha}$ and $u_j^{\beta}$ denote the displacements of atoms $i$ and $j$ from their equilibrium positions in the direction of $\alpha,\beta \in \{x,y,z\}$, $U$ denotes the total potential energy. Although the force constant matrix calculations are limited to second-order, the spectral thermal conductivity is still able to implicitly include high-order phonon-phonon interactions due to the use of fully anharmonic interatomic potentials in the NEMD simulations[53-57]. The final $q(\omega)$ through the virtual interface is expressed by,

$$q(\omega) = \frac{1}{A} \sum_{i \in \tilde{L}} \sum_{j \in \tilde{R}} q_{i \to j}(\omega), \tag{12}$$

where $A$ is the virtual interface area, $\tilde{L}$ and $\tilde{R}$ denote the region within cutoff radius (12Å) on the left and right sides of the virtual interface, respectively. Based on the SHC ($q(\omega)$), the transmission coefficient ($T(\omega)$) can be calculated by,

$$T(\omega) = \frac{q(\omega)}{k_B \Delta T}, \tag{13}$$

where $k_B$ is the Boltzmann constant, $\Delta T$ denotes the temperature difference between the heat source and heat sink. Based on the SHC ($q(\omega)$), the spectral thermal conductivity (STC) is given by,

$$\kappa(\omega) = \frac{q(\omega)}{A \Delta T} L, \tag{14}$$

where $\Delta T$ and $L$ denote the temperature difference between the heat source and heat sink and the corresponding length, respectively. The thermal conductivity of the simulated system can also be calculated from the integral of the STC with frequency,

$$\kappa = \int_{-\infty}^{+\infty} k(\omega) d\omega. \tag{15}$$

## 3. RESULTS AND DISCUSSION

### 3.1. Pair distribution function and angular distribution function analysis.

The pair distribution function (PDF) and angular distribution function (ADF) can be used to characterize the detailed information of the geopolymer structure, and are very common characterization method in both experiment and simulation. Figure 3 shows the PDF of NAS2 and NASH7.5 systems. In addition, the geopolymer PDF measured by White et al[58]. using X-ray diffraction experiment is also plotted in Figure 3 for comparison. It can be seen that the first peak is at the position of 1.06 Å in the PDF of NASH7.5 system, corresponding to the bond length of H-O bond, indicating that there are H-O interaction. The second and third sharp peaks are at the positions of 1.62 Å and 1.78 Å, indicating that there are Si-O interaction and Al-O interaction, respectively. The two peaks obtained here are well correlated with NAS2 system, while only one peak appears at the corresponding position in the experimental measurement by White et al.[58] In fact, it is pointed out in the literature that the bond length of Al-O bond is slightly longer than that of Si-O bond, but it may be limited by the accuracy of experimental measurement, so that no obvious double peaks appear. However, from the interval distribution of these two peaks, there is still a good correlation between the simulated data and the experimental data. The next three peaks are at the positions of 2.4Å, 2.63Å and 3.15Å, representing Na-O, O-O and T-T, T-Na (T represents Si and Al atoms in the silicon tetrahedron and aluminum tetrahedron) interactions, respectively. From the distribution of PDF, it can be seen that the simulated data of the constructed systems are in good agreement with the experimental data, indicating the rationality of the structure and the accuracy of the interatomic potential function used.

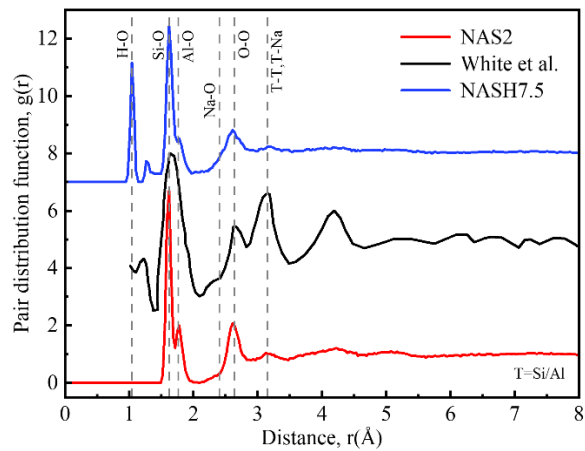

**Figure 3.** Pair distribution function for NAS2 and NASH7.5 systems

Figure 4a and Figure 4b show the ADF of O-Si-O and O-Al-O in NAS2 and NASH7.5 systems, respectively. In addition, the results simulated by using the ReaxFF potential function[30] are also plotted in Figure 4 for comparison. It can be seen from Figure 4a that the peak values of the O-Si-O bond angles in NAS2 and NASH7.5 systems are 109° and 107°, respectively, and the peak value of the reported simulation results is about 109.3°. The peak and interval distributions of these three results are relatively close, indicating that the silicon tetrahedral distribution in the system is reasonable, and that water molecules do not have a significant impact on the O-Si-O bond angles within the silicon tetrahedron. It can be seen from Figure 4b that the O-Al-O bond angle peaks in NAS2 and NASH7.5 systems are 107° and 105°, respectively. However, the O-Al-O distribution in the simulated system of this paper is relatively concentrated (80°-140°), while the O-Al-O distribution range in the simulated system of Lyngdoh et al.[30] is relatively scattered (50°-180°), which may be related to the different potential function used.

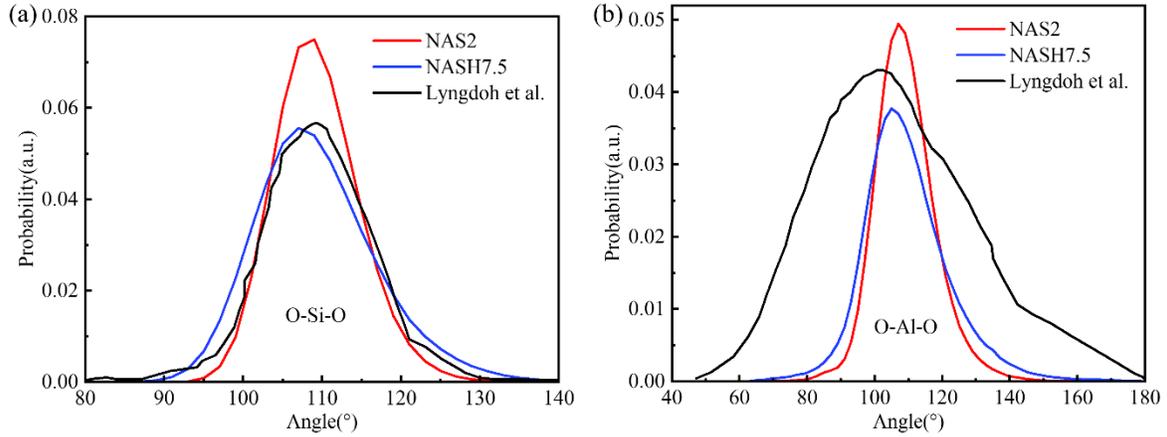

**Figure 4.** (a) ADF of O-Si-O for NAS2 and NASH7.5 systems. (b) ADF of O-Al-O for NAS2 and NASH7.5 systems.

### 3.2. The effect of Si/Al ratio on the thermal conductivity.

Three atomic models (NAS1, NAS2, and NAS3) were first constructed to study the effect of different Si/Al ratio on the thermal conductivity of geopolymer system. According to the thermal conductivity calculation method introduced in the section 2.3.1, the thermal conductivity of all systems are calculated as shown in Figure 5. It can be found that when the Si/Al ratio gradually increases from 1 to 3, the thermal conductivity of the system is 1.665, 1.846, and 1.916 W/(m·K), showing an increasing trend.

In order to investigate the intrinsic mechanism, we further calculated the STC (Figure 6a), the phonon density of states (Figure 6b), the phonon participation ratio (Figure 6c) and the corresponding accumulated thermal conductivity (Figure 6d) curves of different systems, respectively. The STC curve reflects the contribution of different frequency phonons to the thermal conductivity. It can be seen from Figure 6a that the STC of each system is almost identical in the frequency range of 0-2.6 THz and 25-30 THz as the Si/Al ratio increases, which indicates that the phonons contribute less to the increase of thermal conductivity in this frequency region. In the frequency range of 2.6-8.0 THz, there is an overall increasing trend in the STC of each system as the Si/Al ratio increases, but a crossover point appears around 7.0 THz. In the frequency range of 8-25 THz, the STC of each system tends to increase significantly as the Si/Al ratio increases, which indicates that the phonons in this region play a decisive role in the increase of thermal conductivity for different systems.

The PDOS and PPR are effective methods to analyze phonon activity, which will be further analyzed from these two aspects. It can be seen from Figure 6b that with the increase of Si/Al ratio, the PDOS of each system is almost identical in the frequency range of 0-2.6 THz, and the PDOS of each system tends to decrease in the frequency range of 2.6-8.0 THz, while the PDOS of each system changes little in the frequency range of 8-30 THz. In the frequency range of 30-38 THz, when Si/Al ratio increases from 1 to 2, the PDOS increases obviously, while when the Si/Al ratio continues to increase from 2 to 3, the PDOS increases slightly. Generally, there are two modes of phonon localization and delocalization, and the delocalization mode plays an important role in heat transport. The degree of phonon localization can be seen from the PPR curve. Here, 0.15 is used as the dividing line between phonon delocalization and localization.

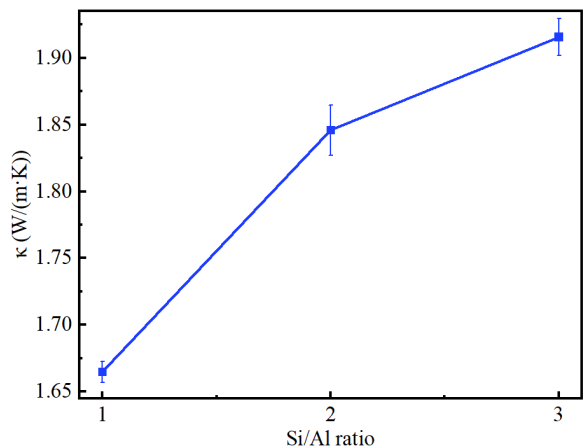

**Figure 5.** The thermal conductivity of different Si/Al ratio systems

It can be seen from Figure 6c that the PPR of geopolymer is generally low, and the highest value is only about 0.5, which reflects the reason why geopolymer has low thermal conductivity to some extent. Relatively speaking, the frequency range with high degree of phonon localization is mainly concentrated in the frequency range of 0-6 THz and 23-40 THz, which indicates that the phonon thermal conduction is seriously blocked in these two frequency intervals, which is not conducive to heat transfer. However, this doesn't necessarily mean these modes can't make significant contributions under certain conditions. For example, the STC curve for the NAS2 system shows a peak value around 3.82 THz, as seen in Figure 6a. Concurrently, the PDOS curve reaches peak value around 3.81 THz (as shown in Figure 6c), suggesting a large number of vibrational modes near this frequency. Even though the PPR value is low, the high PDOS can significantly enhance their total contribution to the thermal conductivity. In the frequency range of 6-23 THz, the phonon participation rate is more than 0.15, and more phonons enter the delocalization state from local to delocalized, which promotes the thermal transportation. In addition, it can be seen from the PDOS distribution of each atom that in the range of 0-8 THz frequency, the contribution of the overall PDOS mainly comes from sodium ions. Since the atomic ratio of Na/Al is always 1, the percentage content of sodium ions will decrease with the increase of Si/Al ratio, which eventually leads to the decrease of PDOS in the low frequency region (0-8 THz). However, sodium ions have little effect on the heat transport performance of geopolymer system, and its heat transport mainly depends on the three-dimensional network structure composed of aluminum tetrahedron and silicon tetrahedron. In the frequency range of 8-30 THz, the PDOS is the result of the interaction of Al, Si and O atoms. When the frequency is higher than 30 THz, the PDOS of both Al atoms and sodium ions are almost zero, and the PDOS is mainly the result of the interaction of Si and O atoms. With the increase of Si/Al ratio, the percentage of Si will increase, and the PDOS in this region will also increase. In addition, from the STC curve, we can see that the cutoff frequency is about 35 THz, which means that when the phonon vibration frequency is higher than 35 THz, there is almost no phonon contribution to the thermal conductivity. However, from the PDOS curve, it can be seen that the cutoff frequency is about 38 THz, which indicates that there is still phonon vibration in the frequency range of 35-38 THz, but it does not have a significant effect on heat transport. The main reason is that in this frequency range, the PPR is almost zero, and the phonon localization is very high, which seriously affects the thermal transport.

It is worth noting that in Figure 6c the geopolymers show lower PPR at lower frequencies. This behavior in amorphous geopolymers is indeed different with acoustic phonon modes in crystals, which are typically expected to have a high PPR near 1. However, our observations align with findings reported in complex amorphous materials like silica glass, as detailed in the provided reference[59]. In amorphous materials, the low-frequency vibrational modes may exhibit atypical behavior due to their unique structural properties, leading to a lower PPR than expected. This suggests that the phonon behavior in our geopolymer study, similar to that in silica glass, may be influenced by specific structural characteristics of the material, resulting in the unusual PPR distribution we observed.

According to the method introduced in the section 2.3.4, the thermal conductivity of the system can be calculated from another point of view. It can be compared with the thermal conductivity calculated by the non-equilibrium molecular dynamics (NEMD) method to verify the correctness of the calculation process. Figure 6d shows the accumulated thermal conductivity curve of different Si/Al ratio systems. We can see from the figure that the accumulated thermal conductivities calculated by this method are 1.617, 1.780, 1.871 W/(m·K), and the errors of these two methods are 2.88%, 3.58%, 2.35%, respectively. The errors are kept below 5%, which proves the accuracy of all the calculation results to a certain extent.

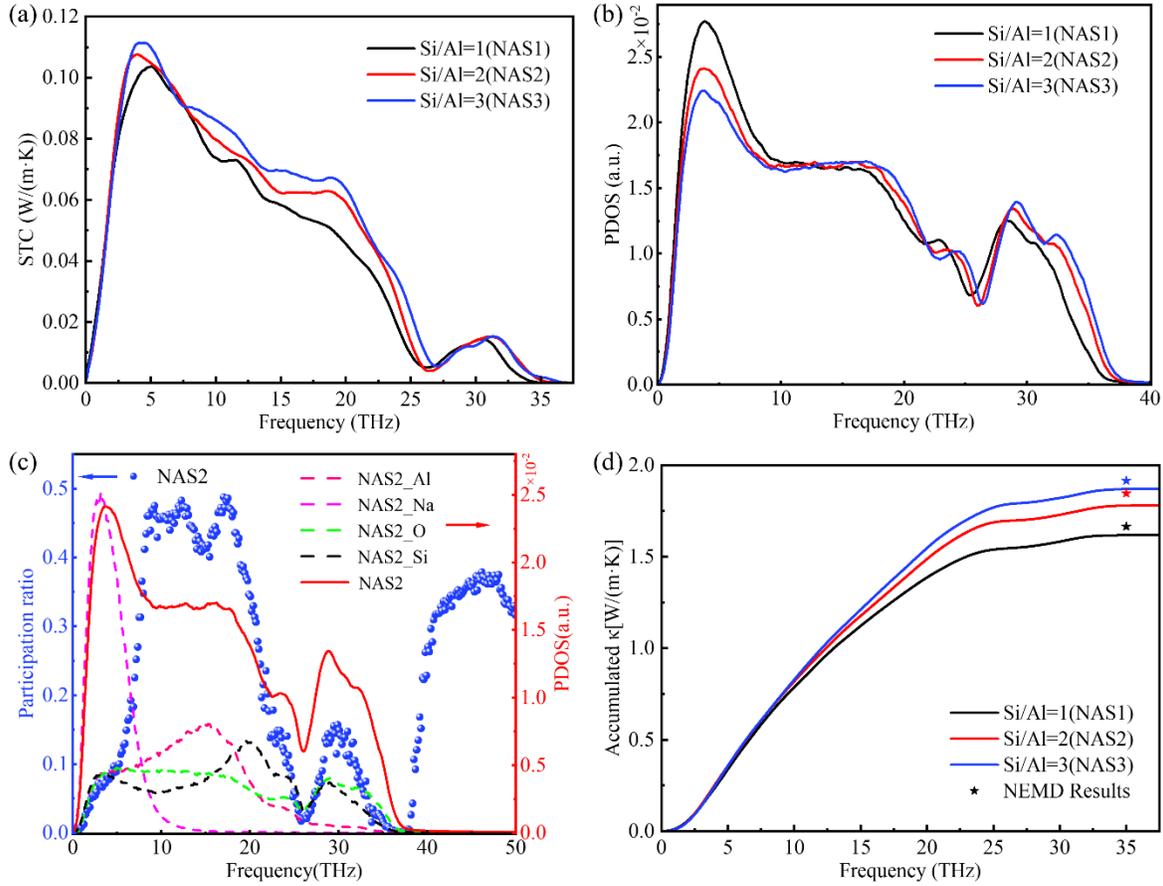

**Figure 6.** (a) The STC curves of different Si/Al ratio systems. (b) The PDOS curves of different Si/Al ratio systems. (c) The PPR curve of NAS2 system and PDOS curves of different atoms in the NAS2 system. (d) The accumulated thermal conductivity curves of different Si/Al ratio systems.

### 3.3. The effect of moisture content on the thermal conductivity.

In order to study the effect and the mechanism of moisture content on the thermal conductivity of geopolymer system. In this work, seven models with moisture content of 0%, 1.25%, 2.50%, 3.75%, 5.00%, 6.25% and 7.50% are respectively constructed, and their serial numbers are NAS2, NASH1.25, NASH2.5, NASH3.75, NASH5, NASH6.25 and NASH7.5. According to the thermal conductivity calculation method introduced in the section 2.3.1., the thermal conductivity of all systems can be calculated (as shown in Figure 7a). It can be found that when the content of moisture increases gradually from 0% to 7.5%, the thermal conductivity of the system is 1.846, 1.641, 1.323, 1.192, 1.103, 1.147, 1.148 W/(m·K), respectively. There is a tendency to decrease at first and then increase and then stabilize at last.

There are many literature studies [60-62] show that the thermal conductivity of a material has a great relationship with its porosity. Generally speaking, if the porosity of the material is higher, it will lead to higher phonon scattering in the process of heat transfer, which is not conducive to the heat transport, thus reducing the thermal conductivity of the material. On the contrary, if the porosity is lower, the phonon scattering effect will be smaller, so that the thermal conductivity of the material will be higher. The porosity and pore size distribution (PSD) of each system are calculated by the Porosity Plus software[63], and the results are shown in Figure 7b and Figure 8, respectively.

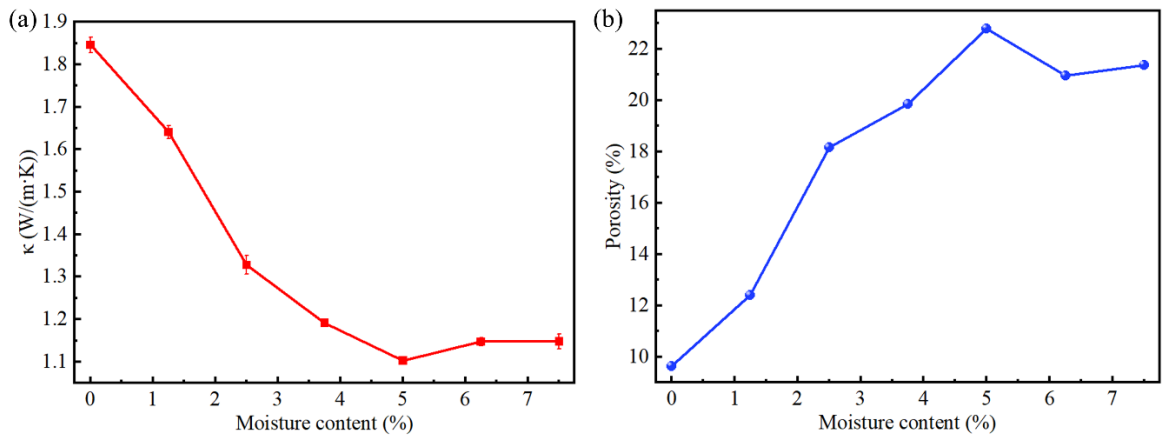

**Figure 7.** (a) The thermal conductivity of different moisture content systems. (b) The porosity curve of different moisture content systems.

From the porosity curve in Figure 7b, it can be seen that with the increase of moisture content, the porosity of the material is 9.64%, 12.39%, 18.16%, 19.85%, 22.80%, 20.95% and 21.38%, which tends to increase at first and then decrease and finally tends to be stable. This is contrary to the calculation results of thermal conductivity, which also shows that for the water-containing geopolymer systems, the thermal conductivity is indeed closely related to porosity. In addition, it can also be seen from the PSD curve in Figure 8 that with the increase of moisture content, the number of the same diameter pores and the size of overall pores diameter in the system tend to increase, and when the moisture content increases to 5%, these two values reach the maximum, and when the moisture content continues to increase, these two values tend to decrease, which will be analyzed in the following discussion. In addition, we can find that when the content of moisture gradually increases from 0% to 1.25%, there are basically only small pores with a diameter of less than 3 Å in the system. When the content of moisture increases to 2.5%, medium pores with diameters of 3-5 Å gradually appear in the system. When the content of moisture continues to increase, there will appear macro pores in the system. According to the simulation results, the maximum pore diameter in the system is about 6 Å. In order to further understand how the pores are distributed in the system with different moisture content, we use OVITO to draw the three-dimensional pore size distribution of NAS2, NASH2.5, NASH5 and NASH7.5 systems, as shown in Figure 9. Where the red area represents pores with a diameter less than 3 Å, and the blue area represents pores with a diameter greater than 3 Å.

It can be seen from Figure 9a that when there are no water molecules in the system, the pores are relatively small and the pore distribution is uniform. In this case, the thermal resistance is relatively small, which is more conducive to heat transport, so the thermal conductivity of the system is the highest. When the moisture content in the system increases to 2.5% (as shown in Figure 9b), it can be observed that the pores significantly increase, and pores with diameters larger than 3 Å appear. These pores are no longer uniformly distributed but exhibit pore clustering. This is due to the fact that water molecules in the system mainly exist in the form of clusters[64], which destroys the integrity of the tetrahedral network. The interaction between water molecule clusters and the tetrahedral network is mainly weak van der Waals forces. As a result, there are many aggregated pores near water molecule clusters, and these pores further increases the interfacial thermal resistance at the interface between water molecule clusters and the tetrahedral network. When heat is transferred from silicon and aluminum tetrahedron to water molecule clusters or from water molecule clusters to silicon and aluminum tetrahedron, there will be obvious phonon scattering at the interface, which will prevent heat transport and reduce the thermal conductivity of the system. When the moisture content in the system increases to 5% (as shown in Figure 9c), the phenomenon of pores aggregation becomes more obvious. However, when the moisture content in the system increases to 7.5% (as shown in Figure 9d), it can be seen that the number of pores do not continue to increase, but decrease slightly. This is mainly because the water molecule clusters have reached saturation, when the moisture content continues to increase in the finite volume, the adjacent water molecule clusters will merge with each other, and the small water molecule clusters will become larger water molecule clusters, thus reducing the interface area between the water molecule clusters and the silicon/aluminum tetrahedra, and finally reducing the porosity of the system. Therefore, the porosity of the system is influenced by both the number and size of water molecule clusters, and when the two factors reach equilibrium, the porosity of the system will become stable. In addition, recent studies have shown that there is a complex interaction between water molecules and thermal conduction mechanisms[65-66]. Water molecules can scatter phonons, reducing thermal conductivity, but can also act as an additional pathway for thermal energy exchange, potentially increasing thermal conductivity. The overall impact on thermal conductivity results from the balance between these opposing effects. Further investigation of the specific role of moisture in geopolymers could enrich our understanding of their thermal properties. This may be an important direction for future research, providing a more detailed perspective on the thermal conduction mechanisms in these materials.

In micro/nanoscale heat transfer, the properties of phonons play an important role in heat transport, because the heat transfer is mainly carried out by phonon vibration. The STC, PDOS, and PPR curves contain different information of phonons, which are commonly used methods to analyze phonon activity. Therefore, we will analyze the microscopic heat transfer mechanism from perspective of phonon vibration.

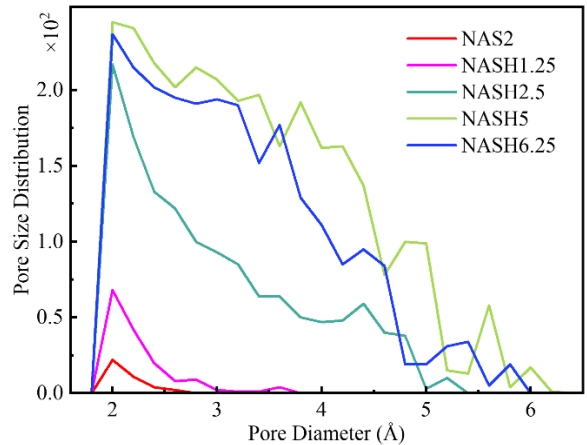

**Figure 8.** The pore size distribution curves of different moisture content systems.

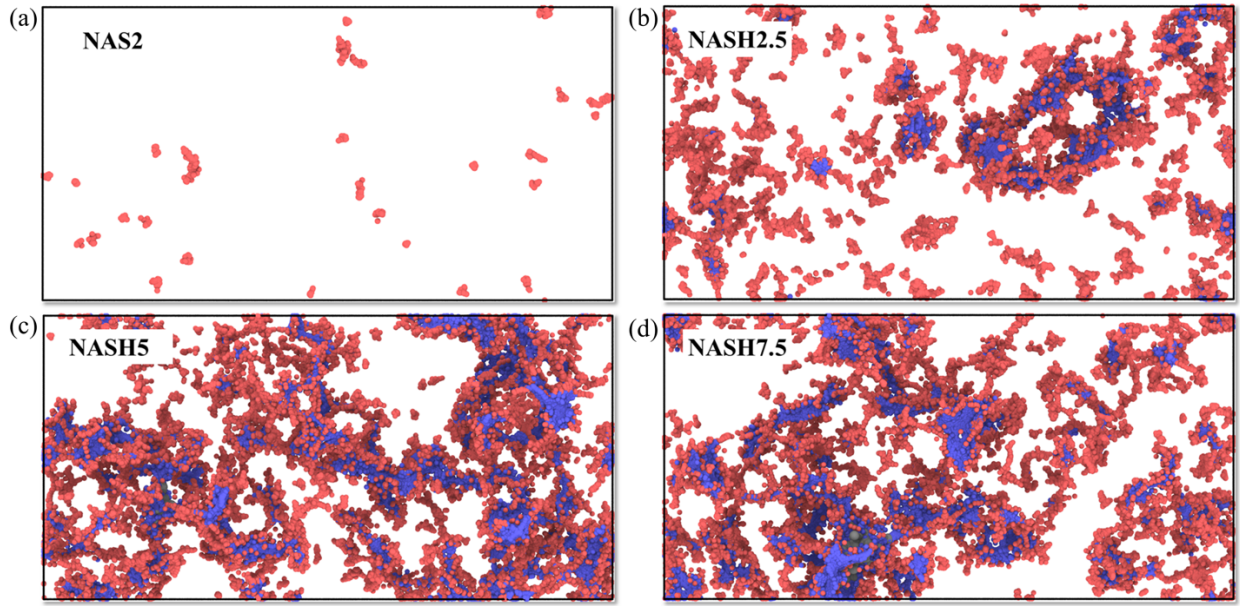

**Figure 9.** (a-d) Three-dimensional pore size distribution of different moisture content systems.

Figure 10a shows the STC curves of different moisture content systems. For clarity, two data sets (NASH1.25 and NASH6.25) that have minimal impact on the analysis result are omitted here. It can be seen from the diagram that when the moisture content increases from 0% to 2.5%, the STC of the system remains almost unchanged in the frequency range of 0-1.6 THz, while the STC increases slightly in the frequency range of 24.7-28.4 THz. However, in the frequency ranges of 1.6-24.7 THz and 28.4-35 THz, the STC of the system decreases obviously. This is also the phonon vibration frequency range which plays an important role in reducing the final thermal conductivity of the system. When the moisture content increases from 2.5% to 5%, the STC remains almost unchanged in the frequency range of 0-1.6 THz and 22-35 THz, and only a few frequencies decrease slightly, while in the frequency range of 1.6-22 THz, the STC of the system has an obvious decreasing trend. When the moisture content increases to 7.5%, the STC of the system increases as a whole, but the trend is not obvious, which indicates that the thermal conductivity of the system with 7.5% moisture content is only slightly higher than that of 5% moisture content system. This is consistent with our previous analysis results.

Figure 10b shows the variation curve of PDOS with frequency for different moisture content systems. It can be seen from the diagram that with the increase of moisture content, the PDOS of each system remains almost unchanged in the frequency range of 0-1.6 THz, and the PDOS decreases as a whole in the frequency range of 1.6-20 THz. In the frequency range of 20-28.4 THz, the overall PDOS tends to increase first, then decrease, and finally stabilize. In addition, when the moisture content is 5%, it reaches the maximum value, which is opposite to the thermal conductivity. In the frequency range of 28.4-35THz, the PDOS tends to decrease as a whole. When the phonon vibration frequency is greater than 35THz, two obvious peaks can be observed, which appear in the frequency range of 40-50THz and 80-93THz, respectively. However, these two peaks only exist in the system containing water molecules, while in the system without water molecules (NAS2), the PDOS reaches its cutoff frequency at 40THz, which indicates that the appearance of the two peaks is caused by the addition of water molecules.

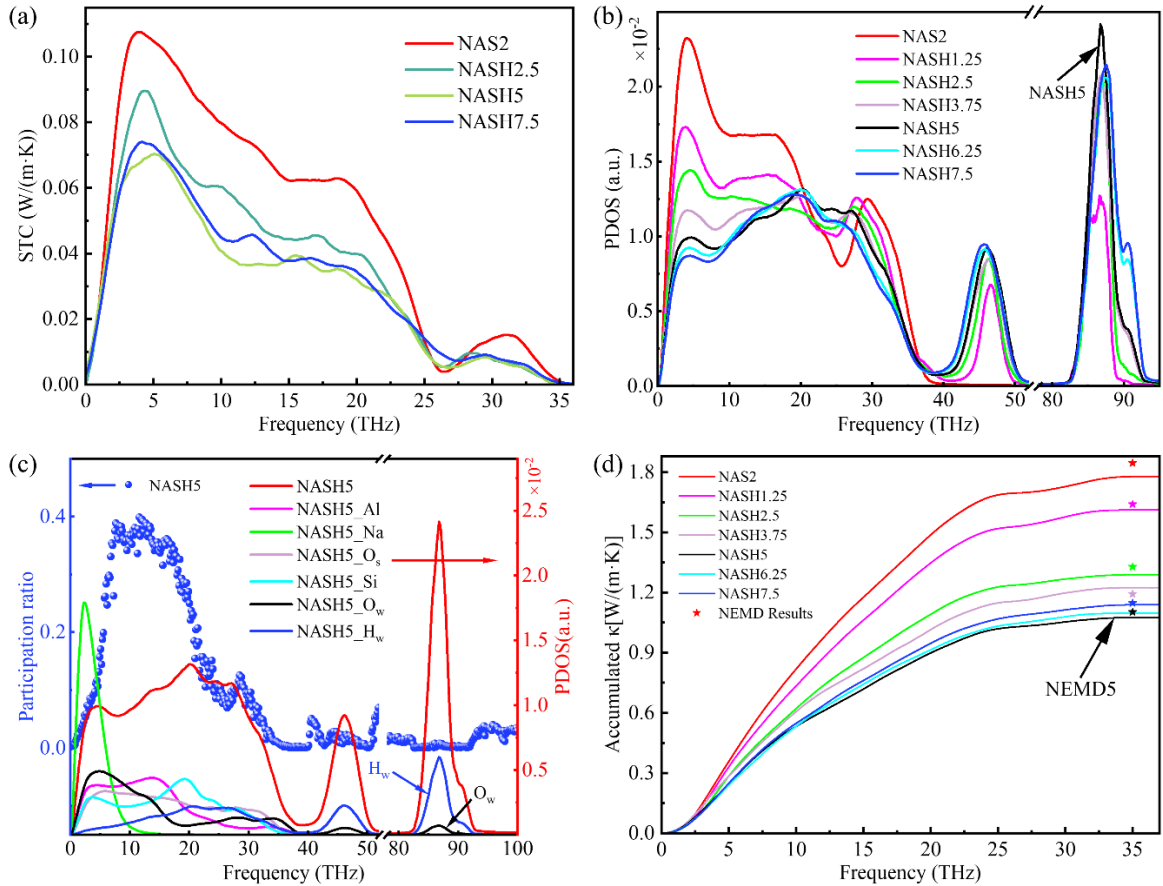

**Figure 10.** (a) The STC curves of different moisture content systems. (b) The PDOS curves of different moisture content systems. (c) The PPR curve of NASH5 system and PDOS curves of different atoms in the NASH5 system. (d) The accumulated thermal conductivity curves of different moisture content systems.

Figure 10c is the variation curve of PPR and PDOS of each atom with frequency for the 5% moisture content system (NASH5). It can be seen from the diagram that the PPR of the system is basically less than 0.1 in the frequency range of 0-4 THz and higher than 23 THz, indicating that the degree of phonon localization is relatively high, which is not conducive to heat transport. In addition, in the frequency range of 40-50 THz and 80-90 THz, only oxygen and hydrogen atoms in water molecules have PDOS distribution, which also causes an abrupt change of the PDOS distribution when water molecules are added to the system.

According to the analysis results of STC, PDOS and PPR, we can find that the PDOS decreases with the increase of moisture content in the frequency range of 1.6-20 THz, while the STC has a minimum value in 5% moisture content system, and the two trends are not consistent. This is mainly because the addition of water molecule leads to the increase of interfacial thermal resistance. When the moisture content is higher than a certain degree, the effect of interfacial thermal resistance on the thermal conductivity is greater than that of PDOS. Therefore, when the moisture content is 5%, the interfacial thermal resistance between water molecule and aluminum/silicon tetrahedron reaches the maximum. When the moisture content continues to increase, these water molecules can start to form continuous pathways. This could facilitate phonon transport and decrease the overall interface thermal resistance, potentially leading to a rising trend in the STC of the system. In addition, in the frequency range of 40-50 THz and 80-93 THz, the water-containing systems have high PDOS, but the STC in this frequency range are almost zero. This is mainly because the PPR in these two frequency ranges is almost zero, which means that the phonon localization in this region is so high that it has almost no effect on heat transport. Therefore, the final thermal conductivity of the system is actually affected by the interfacial thermal resistance, PDOS and PPR.

The above-mentioned rules on the influence of moisture content on the interface thermal resistance are only qualitative analysis and may not be convincing. Next, we will analyze the influence of moisture content on the interface thermal resistance from a computational perspective. When considering interfacial thermal conduction, the matching degree of PDOS is crucial to the transfer efficiency of heat flow. If the PDOS curves of two materials overlap to a high degree, it means that their phonon patterns at the interface match well, and phonons can be transferred from one material to the other more easily, thereby reducing the interface thermal resistance and enhancing thermal conduction. On the contrary, if the PDOS curves overlap area decrease, it indicates that the phonon mode matching degree decreases, resulting in increased scattering of phonons at the interface, thereby increasing the interface thermal resistance and reducing thermal conduction efficiency. Therefore, we can use the overlapping area of the PDOS of geopolymer matrix and the PDOS of water molecules to characterize the size of the interface thermal resistance[49]. Its expression is:

$$S = \int_0^\infty \min \{P_{NAS}(\omega), P_{H_2O}(\omega)\} d\omega. \tag{16}$$

Where $P_{NAS}(\omega)$ and $P_{H_2O}(\omega)$ are the two PDOS of geopolymer matrix and water molecules, respectively. Figure 11a-d clearly shows that as the moisture content increases, the overlap area $S$ shows a trend of first decreasing and then increasing. And when the moisture content is 5%, the overlapping area $S$ reaches the minimum value, which means that the interface thermal resistance in the system reaches the maximum value.

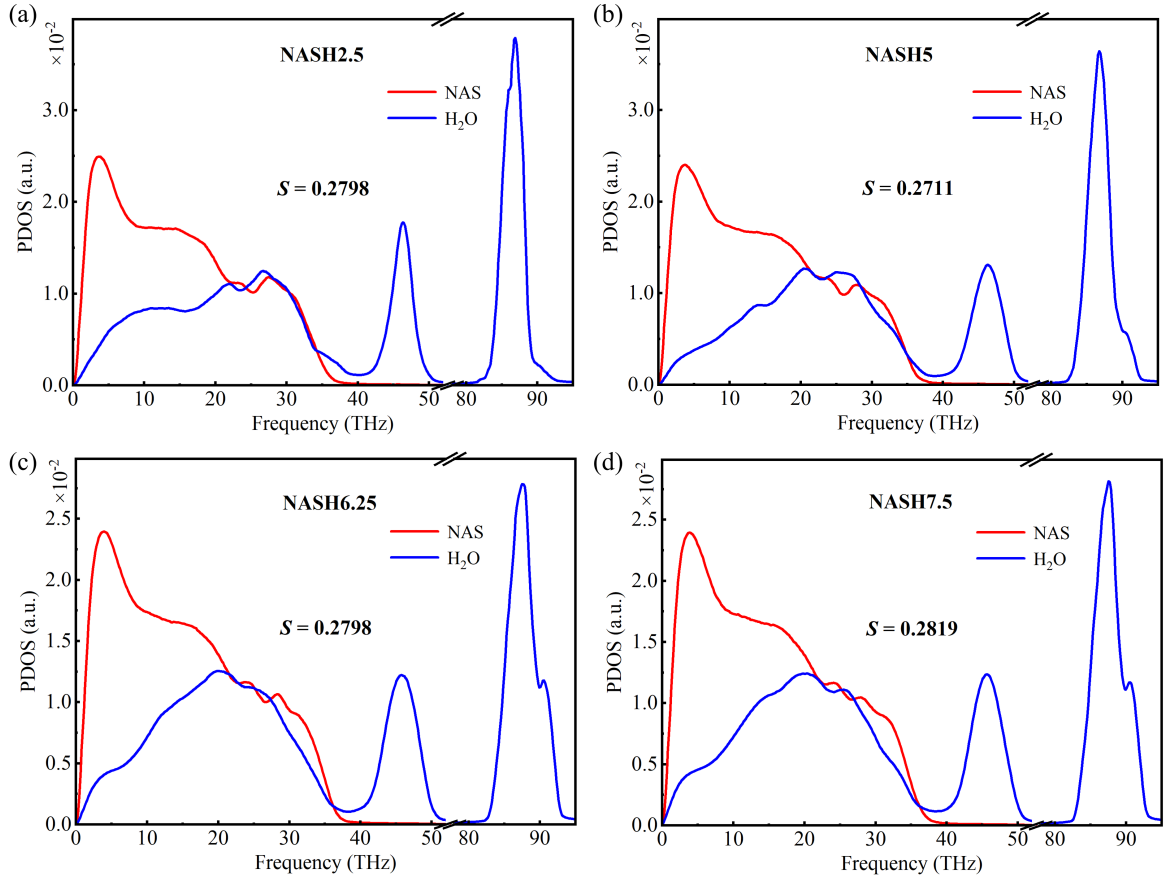

**Figure 11.** (a-d) The PDOS curves of Geopolymer matrix (NAS) and water molecules (H2O) for different moisture content systems.

According to the method introduced in the section 2.3.4, the thermal conductivity of the system can be calculated from another point of view. This can be compared with the thermal conductivity calculated by the NEMD method to verify the correctness of the calculation process. Figure 10d shows the accumulated thermal conductivity curves of different moisture content systems. It can be seen from the figure that the thermal conductivity calculated by this method is 1.780 1.612, 1.290, 1.224, 1.076, 1.098 and 1.140 W/(m·K), respectively, and the error with the NEMD method is 3.59%, 1.78%, 1.89%, 2.72%, 2.52%, 4.29% and 0.71%, respectively. All the errors are kept below 5%, which proves the accuracy of the whole calculation result to a certain extent.

## 4. CONCLUSIONS

In conclusion, the thermal conductivity of geopolymer systems with different Si/Al ratio and moisture content has been systematically investigated by using the molecular dynamics simulation. By combining with the analysis of phonon density of state, phonon participation rate and thermal conductivity spectrum decomposition, the microscopic thermal conductivity mechanism is deeply discussed from different view of points to provide a comprehensive understanding of the microscale heat transfer mechanism in the system.

For different Si/Al ratio systems, the thermal conductivity tends to increase with the increase of Si/Al ratio. The phonon vibration frequency region of 8-25 THz plays a major role in increasing the thermal conductivity, while the remaining frequency range has almost no effect on the increase of thermal conductivity. The heat transport mainly depends on the three-dimensional network structure composed of aluminum tetrahedron and silicon tetrahedron. In addition, the PPR of geopolymer materials is relatively low, and the highest value is only about 0.5, which reflects the reason why geopolymers have low thermal conductivity to some extent.

For the water-containing geopolymer systems, with the increase of the moisture content, the thermal conductivity tends to decrease first, then increase, and finally tends to be stable, which is just opposite to the change trend of the porosity. The main reason is that the existence of pores will lead to phonon scattering during heat transfer at the interface, which will affect the thermal conductivity of the material. When the moisture content is 5%, the thermal conductivity reaches the minimum value, about 1.103 W/(m·K), which is 40.2% lower than that of pure geopolymer system. The PPR of the water-containing geopolymer systems in the frequency range of 0-4 THz and higher than 23 THz is basically less than 0.1, indicating that the degree of phonon localization in these frequency regions is high, which is not conducive to heat transfer. The addition of water molecules makes the PDOS produce two peaks in the frequency range of 40-50 THz and 80-90 THz, respectively. However, the STC analysis shows that there is no component in these frequency regions, which is mainly due to the PPR in these regions is almost zero.

This work provides some insights for the study of microscale heat transport properties of geopolymer systems, and has certain guiding significance for future research on geopolymer composite systems and multi-scale simulations.

## AUTHOR INFORMATION

**Corresponding Author**

**Shenghong Ju** - China-UK Low Carbon College, Shanghai Jiao Tong University, Shanghai, 201306, China; Email: shenghong.ju@sjtu.edu.cn

**Author**

**Wenkai Liu** - China-UK Low Carbon College, Shanghai Jiao Tong University, Shanghai, 201306, China

**Notes**

The authors declare no competing financial interest.

## ACKNOWLEDGMENT

This work was supported by the National Natural Science Foundation of China (No. 52006134) and Shanghai Key Fundamental Research Grant (No. 21JC1403300). The computations in this paper were run on the $\pi$ 2.0 cluster supported by the Center for High Performance Computing at Shanghai Jiao Tong University.

REFERENCES

(1) Singh, N. B.; Middendorf, B., Geopolymers as an Alternative to Portland Cement: An Overview. *Construction and Building Materials* **2020**, *237*, 117455.

(2) Al Bakri, A. M.; Kamarudin, H.; Bnhussain, M.; Nizar, I. K.; Mastura, W., Mechanism and Chemical Reaction of Fly Ash Geopolymer Cement-a Review. *Journal of Asian Scientific Research* **2011**, *1*, 247-253.

(3) Wu, Y.; Lu, B.; Bai, T.; Wang, H.; Du, F.; Zhang, Y.; Cai, L.; Jiang, C.; Wang, W., Geopolymer, Green Alkali Activated Cementitious Material: Synthesis, Applications and Challenges. *Construction and Building Materials* **2019**, *224*, 930-949.

(4) Puligilla, S.; Mondal, P., Role of Slag in Microstructural Development and Hardening of Fly Ash-Slag Geopolymer. *Cement and Concrete Research* **2013**, *43*, 70-80.

(5) Molino, B.; De Vincenzo, A.; Ferone, C.; Messina, F.; Colangelo, F.; Cioffi, R., Recycling of Clay Sediments for Geopolymer Binder Production. A New Perspective for Reservoir Management in the Framework of Italian Legislation: The Occhito Reservoir Case Study. *Materials* **2014**, *7*, 5603-5616.

(6) Gopalakrishna, B.; Dinakar, P., Mix Design Development of Fly Ash-Ggbs Based Recycled Aggregate Geopolymer Concrete. *Journal of Building Engineering* **2023**, *63*, 105551.

(7) Zhao, C.; Ju, S.; Xue, Y.; Ren, T.; Ji, Y.; Chen, X., China's Energy Transitions for Carbon Neutrality: Challenges and Opportunities. *Carbon Neutrality* **2022**, *1*, 7.

(8) van Deventer, J. S.; Provis, J. L.; Duxson, P.; Brice, D. G., Chemical Research and Climate Change as Drivers in the Commercial Adoption of Alkali Activated Materials. *Waste and Biomass Valorization* **2010**, *1*, 145-155.

(9) Davidovits, J., False Values on Co2 Emission for Geopolymer Cement/Concrete Published in Scientific Papers. *Technical paper* **2015**, *24*, 1-9.

(10) Zannerni, G. M.; Fattah, K. P.; Al-Tamimi, A. K., Ambient-Cured Geopolymer Concrete with Single Alkali Activator. *Sustainable materials and technologies* **2020**, *23*, e00131.

(11) Peng, C.; Wei, Q.; Wei, W., Decarbonization Path of China's Public Building Sector from Bottom to Top. *Carbon Neutrality* **2022**, *1*, 39.

(12) Luhar, S.; Nicolaides, D.; Luhar, I., Fire Resistance Behaviour of Geopolymer Concrete: An Overview. *Buildings* **2021**, *11*, 82.

(13) Vellattu Chola, R. K.; Ozhukka Parambil, F.; Panakkal, T.; Meethale Chelaveettil, B.; Kumari, P.; Valiya Peedikkakal, S., Clean Technology for Sustainable Development by Geopolymer Materials. *Physical Sciences Reviews* **2022**, *7*, 100445.

(14) Cong, P.; Cheng, Y., Advances in Geopolymer Materials: A Comprehensive Review. *Journal of Traffic and Transportation Engineering (English Edition)* **2021**, *8*, 283-314.

(15) Duxson, P.; Provis, J. L.; Lukey, G. C.; Van Deventer, J. S., The Role of Inorganic Polymer Technology in the Development of 'Green Concrete'. *Cement and Concrete Research* **2007**, *37*, 1590-1597.

(16) Zhang, Z.; Provis, J. L.; Reid, A.; Wang, H., Geopolymer Foam Concrete: An Emerging Material for Sustainable Construction. *Construction and Building Materials* **2014**, *56*, 113-127.

(17) Majidi, B., Geopolymer Technology, from Fundamentals to Advanced Applications: A Review. *Materials Technology* **2009**, *24*, 79-87.

(18) Lyu, S. J.; Wang, T. T.; Cheng, T. W.; Ueng, T. H., Main Factors Affecting Mechanical Characteristics of Geopolymer Revealed by Experimental Design and Associated Statistical Analysis. *Construction and Building Materials* **2013**, *43*, 589-597.

(19) Pradhan, P.; Panda, S.; Parhi, S. K.; Panigrahi, S. K., Factors Affecting Production and Properties of Self-Compacting Geopolymer Concrete–a Review. *Construction and Building Materials* **2022**, *344*, 128174.

(20) Part, W. K.; Ramli, M.; Cheah, C. B., An Overview on the Influence of Various Factors on the Properties of Geopolymer Concrete Derived from Industrial by-Products. *Construction and Building Materials* **2015**, *77*, 370-395.

(21) Davidovits, J., Geopolymers: Inorganic Polymeric New Materials. *Journal of Thermal Analysis and calorimetry* **1991**, *37*, 1633-1656.

(22) Sadat, M. R.; Bringuier, S.; Muralidharan, K.; Runge, K.; Asaduzzaman, A.; Zhang, L., An Atomistic Characterization of the Interplay between Composition, Structure and Mechanical Properties of Amorphous Geopolymer Binders. *Journal of Non-Crystalline Solids* **2016**, *434*, 53-61.

(23) Ahmed, M. M.; El-Naggar, K. A. M.; Tarek, D.; Ragab, A.; Sameh, H.; Zeyad, A. M.; Tayeh, B. A.; Maafa, I. M.; Yousef, A., Fabrication of Thermal Insulation Geopolymer Bricks Using Ferrosilicon Slag and Alumina Waste. *Case Studies in Construction Materials* **2021**, *15*, e00737.

(24) Ranjbar, N.; Mehrali, M.; Alengaram, U. J.; Metselaar, H. S. C.; Jumaat, M. Z., Compressive Strength and Microstructural Analysis of Fly Ash/Palm Oil Fuel Ash Based Geopolymer Mortar under Elevated Temperatures. *Construction and Building Materials* **2014**, *65*, 114-121.

(25) Kljajević, L.; Nenadović, M.; Ivanović, M.; Bučevac, D.; Mirković, M.; Mladenović Nikolić, N.; Nenadović, S., Heat Treatment of Geopolymer Samples Obtained by Varying Concentration of Sodium Hydroxide as Constituent of Alkali Activator. *Gels* **2022**, *8*, 333.

(26) Škvára, F.; Kopecký, L.; Šmilauer, V.; Bittnar, Z., Material and Structural Characterization of Alkali Activated Low-Calcium Brown Coal Fly Ash. *Journal of hazardous materials* **2009**, *168*, 711-720.

(27) Zuhua, Z.; Xiao, Y.; Huajun, Z.; Yue, C., Role of Water in the Synthesis of Calcined Kaolin-Based Geopolymer. *Applied Clay Science* **2009**, *43*, 218-223.

(28) White, C. E.; Provis, J. L.; Proffen, T.; Van Deventer, J. S., The Effects of Temperature on the Local Structure of Metakaolin‐Based Geopolymer Binder: A Neutron Pair Distribution Function Investigation. *Journal of the American Ceramic Society* **2010**, *93*, 3486-3492.

(29) Fang, Y.; Kayali, O., The Fate of Water in Fly Ash-Based Geopolymers. *Construction and Building Materials* **2013**, *39*, 89-94.

(30) Lyngdoh, G. A.; Kumar, R.; Krishnan, N.; Das, S., Realistic Atomic Structure of Fly Ash-Based Geopolymer Gels: Insights from Molecular Dynamics Simulations. *The Journal of Chemical Physics* **2019**, *151*, 064307.

(31) Sadat, M. R.; Bringuier, S.; Asaduzzaman, A.; Muralidharan, K.; Zhang, L., A Molecular Dynamics Study of the Role of Molecular Water on the Structure and Mechanics of Amorphous Geopolymer Binders. *The Journal of Chemical Physics* **2016**, *145*, 134706.

(32) Singh, N. B., Foamed Geopolymer Concrete. *Materials Today: Proceedings* **2018**, *5*, 15243-15252.

(33) Van Gunsteren, W. F.; Berendsen, H. J., Computer Simulation of Molecular Dynamics: Methodology, Applications, and Perspectives in Chemistry. *Angewandte Chemie International Edition in English* **1990**, *29*, 992-1023.

(34) Allen, M. P., Introduction to Molecular Dynamics Simulation. *Computational soft matter: from synthetic polymers to proteins* **2004**, *23*, 1-28.

(35) Lau, D.; Jian, W.; Yu, Z.; Hui, D., Nano-Engineering of Construction Materials Using Molecular Dynamics Simulations: Prospects and Challenges. *Composites Part B: Engineering* **2018**, *143*, 282-291.

(36) Lahoti, M.; Wong, K. K.; Yang, E. H.; Tan, K. H., Effects of Si/Al Molar Ratio on Strength Endurance and Volume Stability of Metakaolin Geopolymers Subject to Elevated Temperature. *Ceramics International* **2018**, *44*, 5726-5734.

(37) Guan, X.; Xu, M.; Li, B.; Do, H., Interactions between Amorphous Silica and Sodium Alumino-Silicate Hydrate Gels: Insight from Reactive Molecular Dynamics Simulation. *The Journal of Physical Chemistry C* **2023**, *127*, 13302-13316.

(38) Liu, W.; Qin, L.; Zhao, C.; Ju, S., Microscopic Mechanism of Tunable Thermal Conductivity in Carbon Nanotube-Geopolymer Nanocomposites. *The Journal of Physical Chemistry B* **2023**, *127*, 2267-2276.

(39) Chanajaree, R.; Bopp, P. A.; Fritzsche, S.; Kärger, J., Water Dynamics in Chabazite. *Microporous and mesoporous materials* **2011**, *146*, 106-118.

(40) Sperinck, S. Metakaolin as a Model System for Understanding Geopolymers. Curtin University, 2012.

(41) Malavasi, G.; Menziani, M. C.; Pedone, A.; Segre, U., Void Size Distribution in Md-Modelled Silica Glass Structures. *Journal of Non-Crystalline Solids* **2006**, *352*, 285-296.

(42) Tilocca, A.; de Leeuw, N. H.; Cormack, A. N., Shell-Model Molecular Dynamics Calculations of Modified Silicate Glasses. *Physical Review B* **2006**, *73*, 104209.

(43) Bauchy, M., Structural, Vibrational, and Thermal Properties of Densified Silicates: Insights from Molecular Dynamics. *The Journal of Chemical Physics* **2012**, *137*, 044510.

(44) Wu, Y.; Tepper, H. L.; Voth, G. A., A Flexible Simple Point-Charge Water Model with Improved Liquid-State Properties. *The Journal of Chemical Physics* **2006**, *124*, 024503.

(45) Plimpton, S., Fast Parallel Algorithms for Short-Range Molecular Dynamics. *Journal of computational physics* **1995**, *117*, 1-19.

(46) Huang, W. X.; Pei, Q. X.; Liu, Z. S.; Zhang, Y. W., Thermal Conductivity of Fluorinated Graphene: A Non-Equilibrium Molecular Dynamics Study. *Chemical Physics Letters* **2012**, *552*, 97-101.

(47) Nosé, S., A Unified Formulation of the Constant Temperature Molecular Dynamics Methods. *The Journal of Chemical Physics* **1984**, *81*, 511-519.

(48) Hoover, W. G., Canonical Dynamics: Equilibrium Phase-Space Distributions. *Physical Review A* **1985**, *31*, 1695.

(49) Liang, T.; Zhou, M.; Zhang, P.; Yuan, P.; Yang, D., Multilayer in-Plane Graphene/Hexagonal Boron Nitride Heterostructures: Insights into the Interfacial Thermal Transport Properties. *International Journal of Heat and Mass Transfer* **2020**, *151*, 119395.

(50) Dickey, J.; Paskin, A., Computer Simulation of the Lattice Dynamics of Solids. *Physical Review* **1969**, *188*, 1407.

(51) Wu, X.; Han, Q., Semidefective Graphene/H-Bn in-Plane Heterostructures: Enhancing Interface Thermal Conductance by Topological Defects. *The Journal of Physical Chemistry C* **2021**, *125*, 2748-2760.

(52) Sääskilahti, K.; Oksanen, J.; Tulkki, J.; Volz, S., Role of Anharmonic Phonon Scattering in the Spectrally Decomposed Thermal Conductance at Planar Interfaces. *Physical Review B* **2014**, *90*, 134312.

(53) Sääskilahti, K.; Oksanen, J.; Volz, S.; Tulkki, J., Frequency-Dependent Phonon Mean Free Path in Carbon Nanotubes from Nonequilibrium Molecular Dynamics. *Physical Review B* **2015**, *91*, 115426.

(54) Zhou, Y.; Hu, M., Full Quantification of Frequency-Dependent Interfacial Thermal Conductance Contributed by Two-and Three-Phonon Scattering Processes from Nonequilibrium Molecular Dynamics Simulations. *Physical Review B* **2017**, *95*, 115313.

(55) Xu, Y. X.; Fan, H. Z.; Zhou, Y. G., Quantifying Spectral Thermal Transport Properties in Framework of Molecular Dynamics Simulations: A Comprehensive Review. *Rare Metals* **2023**, *42*, 3914-3944.

(56) Xu, Y.; Yang, L.; Zhou, Y., The Interfacial Thermal Conductance Spectrum in Nonequilibrium Molecular Dynamics Simulations Considering Anharmonicity, Asymmetry and Quantum Effects. *Physical Chemistry Chemical Physics* **2022**, *24*, 24503-24513.

(57) Xu, Y.; Fan, H.; Li, Z.; Zhou, Y., Signatures of Anharmonic Phonon Transport in Ultrahigh Thermal Conductance across Atomically Sharp Metal/Semiconductor Interface. *International Journal of Heat and Mass Transfer* **2023**, *201*, 123628.

(58) White, C. E.; Page, K.; Henson, N. J.; Provis, J. L., In Situ Synchrotron X-Ray Pair Distribution Function Analysis of the Early Stages of Gel Formation in Metakaolin-Based Geopolymers. *Applied Clay Science* **2013**, *73*, 17-25.

(59) Bonfanti, S.; Guerra, R.; Mondal, C.; Procaccia, I.; Zapperi, S., Universal Low-Frequency Vibrational Modes in Silica Glasses. *Physical Review Letters* **2020**, *125*, 085501.

(60) Liu, H.; Zhao, X., Thermal Conductivity Analysis of High Porosity Structures with Open and Closed Pores. *International Journal of Heat and Mass Transfer* **2022**, *183*, 122089.

(61) Hong, S. N.; Yu, C. J.; Ri, K. C.; Han, J. M.; Ri, B. H., Molecular Dynamics Study of the Effect of Moisture and Porosity on Thermal Conductivity of Tobermorite 14 Å. *International Journal of Thermal Sciences* **2021**, *159*, 106537.

(62) Chi, W.; Sampath, S.; Wang, H., Microstructure‐Thermal Conductivity Relationships for Plasma‐Sprayed Yttria‐Stabilized Zirconia Coatings. *Journal of the American Ceramic Society* **2008**, *91*, 2636-2645.

(63) Opletal, G.; Petersen, T. C.; Russo, S. P.; Barnard, A. S., Porosityplus: Characterisation of Defective, Nanoporous and Amorphous Materials. *Journal of Physics: Materials* **2018**, *1*, 016002.

(64) Barbosa, V. F.; MacKenzie, K. J.; Thaumaturgo, C., Synthesis and Characterisation of Materials Based on Inorganic Polymers of Alumina and Silica: Sodium Polysialate Polymers. *International Journal of Inorganic Materials* **2000**, *2*, 309-317.

(65) Fan, H.; Yang, C.; Zhou, Y., Ultralong Mean Free Path Phonons in Hkust-1 and Their Scattering by Water Adsorbates. *Physical Review B* **2022**, *106*, 085417.

(66) Babaei, H.; DeCoster, M. E.; Jeong, M.; Hassan, Z. M.; Islamoglu, T.; Baumgart, H.; McGaughey, A. J.; Redel, E.; Farha, O. K.; Hopkins, P. E., Observation of Reduced Thermal Conductivity in a Metal-Organic Framework Due to the Presence of Adsorbates. *Nature communications* **2020**, *11*, 4010.